\documentclass[11pt,fleqn]{article}
\usepackage[dvips]{graphicx}
\usepackage{epsf}
\usepackage{latexsym}

\newtheorem{theorem}{Theorem}
\newtheorem{corollary}[theorem]{Corollary}
\newtheorem{definition}[theorem]{Definition}
\newtheorem{lemma}[theorem]{Lemma}

\newtheorem{proposition}[theorem]{Proposition}

\newenvironment{proof}{\noindent{\it Proof. }} {{\qed}}
\newenvironment{proofof}[1]{\noindent{\it Proof of #1. }} {{\qed}}

\def\qed{\hspace*{\fill} $\Box$\par\medskip}

\def\email#1{\tt #1}

\usepackage{amssymb}

\newcommand{\EXP}{\mbox{\rm EXP}}
\newcommand{\BPP}{\mbox{\rm BPP}}

\renewcommand{\P}{{\rm P}}

\newcommand{\NP}{\mbox{\rm NP}}

\newcommand{\REC}{\mbox{\rm REC}}

\newcommand{\SAT}{\mbox{\rm SAT}}

\catcode`\@=11
\let\@originalMarginpar=\marginpar
\newcommand{\Comment}[1]{\@originalMarginpar{\raggedright\footnotesize\em#1}}
\def\marginpar#1{\Comment{#1}}
\catcode`\@=12

\newcommand{\PSPACE}{{\rm{PSPACE}}}

\newlength{\infoboxwidth}
\newcommand{\ioEXP}{\mbox{\rm i.o.-EXP}}
\newcommand{\ioM}{\mbox{\rm i.o.-M}}
\newcommand{\ioU}{\mbox{\rm i.o.-U}}

\newcommand{\ioDTIME}{\mbox{\rm i.o.-DTIME}}

\newcommand{\PTIME}{\mbox{\rm P}}
\newcommand{\Ptime}{\PTIME}

\newcommand{\C}{\mathcal{C}}
\newcommand{\B}{\mathcal{B}}

\def\N{{\mathbf{N}}}

\def\ptt{\mathop{\leq^{\rm p}_{\rm tt}}\nolimits}

\def\pdtt{\mathop{\leq^{\rm p}_{\rm dtt}}\nolimits}
\def\pctt{\mathop{\leq^{\rm p}_{\rm ctt}}\nolimits}

\begin{document}

\title{Derandomizing from Random Strings}

\author{
Harry Buhrman\thanks{Partially supported by an NWO  VICI grant.}\\
CWI and University of Amsterdam\\
\email{buhrman@cwi.nl}\and Lance Fortnow\thanks{Supported in part by
NSF grants CCF-0829754 and DMS-0652521.}\\Northwestern
University\\
\email{fortnow@northwestern.edu}\and
Michal Kouck\'y\thanks{Partially supported by project No.~1M0021620808 of M\v{S}MT \v{C}R and
Institutional Research Plan No.~AV0Z10190503.}\\
Institute of Mathematics, AS CR\\
\email{koucky@math.cas.cz}\and
Bruno Loff\thanks{Supported by
a Portuguese science FCT grant.}\\ CWI\\\email{bruno.loff@gmail.com} }

\date{}

\maketitle

\begin{abstract}
In this paper we show that   BPP is truth-table reducible to the set
of Kolmogorov random strings $R_K$. It was previously known that
PSPACE, and hence BPP is Turing-reducible to $R_K$. The earlier
proof relied on the adaptivity of the Turing-reduction to find a
Kolmogorov-random string of polynomial length using the set $R_K$ as
oracle. Our new non-adaptive result relies on a new fundamental fact
about the set $R_K$, namely each initial segment of the
characteristic sequence of $R_K$ is not compressible by recursive
means. As a partial converse to our claim we show that strings of
high Kolmogorov-complexity when used as advice are not much more
useful than randomly chosen strings.
\end{abstract}

\section{Introduction}
Kolmogorov complexity studies the amount of randomness in a string
by the smallest program that can generate it. The most random
strings are those we cannot compress at all making the set $R_K =
\{x\mathrel{|} K(x) \geq |x|\}$ of Kolmogorov random strings worthy
of close analysis.

%In this paper we continue to explore the question of which sets can
%be efficiently reduced to the set of Kolmogorov random
%strings~\cite{AllenderBurmanKouckyRonneburgMelkebeek06,AllenderBuhrmanKoucky06}.
Allender et al.~\cite{ABKvMR02}
showed the surprising computational power of $R_K$ including that
polynomial time adaptive (Turing) access to $R_K$ enables one to do
PSPACE-computations: $\PSPACE \subseteq P^{R_K}$. One of the
ingredients in the proof shows how on input $0^n$ one can in
polynomial time with {\em adaptive} access to $R_K$ generate a
polynomially long Kolmogorov random string. With {\em non-adaptive}
access it is only possible  to generate in polynomial time a random
string of length at most $O(\log n)$.

In an attempt to {\em characterize} PSPACE as the class of sets
reducible to $R_K$, Allender, Buhrman and
Kouck\'y~\cite{ABK} noticed that this question
depends on the choice of universal machine used in the definition of
the notion of Kolmogorov complexity. They also started a systematic
study of weaker and non-adaptive access to $R_K$. They showed for
example that \[\P = \REC \cap\bigcap_U \{A\mathrel{|} A\pdtt
R_{K_U}\}.\] This result and the fact that with non-adaptive access
to $R_K$ in general only logarithmically small strings can be found
seems to suggest that adaptive access to $R_K$ is needed in order to
be useful.

Our first result proves this intuition false: We show that
polynomial time non-adaptive access to $R_K$ can be used to
derandomize any BPP computation. In order to derandomize a BPP
computation one needs a (pseudo)random string of polynomial size. As
mentioned before one can only obtain short, $O(\log n)$ sized,
random strings from $R_K$. Instead we show that the characteristic
sequence formed by the strings of length $c\log n$, $R_K^{= c\log
n}$, itself a strings of length $n^c$, is complex enough to figure
as a hard function in the hardness versus randomness framework of
Impagliazzo and Wigderson~\cite{IW97}. This way we construct a
pseudorandom generator that is strong enough to derandomize BPP.

In particular we show that for every time bound $t$, there is a
constant $c$ such that  $R_K \not \in \ioDTIME(t)/2^{n-c}$. This is
in stark contrast with the time-unbounded case where only $n$
bits of advice are necessary~\cite{Barzdin}. As a consequence we
give an alternative proof of the existence of an r.e. set $A$, due
to Barzdin~\cite{Barzdin}, such that for all time bounds $t$, there
exists $c_t$ such that $K^{t(n)}(A_{1:n}\mathrel{|}n) \geq n/c_t $. We
simply take for $A$ the complement of  $R_K$. Barzdin also showed
that this lower bound is optimal for r.e. sets. Hence the constant
depending on the time-bound in our Theorem~\ref{t-RK} is optimal.

Next we try to establish whether we can characterize BPP as the
class of  sets that non-adaptively reduce  to $R_K$. One can view
the truth-table reduction to $R_K$ as a computation
with advice of $K^{t(n)}$ complexity $\Omega(n)$. We can show that
for sets in EXP and $t(n)\in 2^{n^{\Omega(1)}}$, polynomial-time computation
with polynomial (exponential, resp.)
size  advice of $K^{t(n)}$ complexity $n-O(\log n)$ ($n-O(\log \log n)$, resp.)
can be simulated by bounded error probabilistic machine with almost
linear size advice. For paddable sets that are complete for
$\NP, \PTIME^{\#\PTIME}, \PSPACE$, or $\EXP$ we do not even need
the linear size advice. Hence, advice of high $K^{t(n)}$ complexity
is no better than a truly random string.

Summarizing our results:

\begin{itemize}
\item For every computable time bound $t$ there is a constant c (depending on $t$) such that $R_K \not
  \in \ioDTIME(t)/2^{n-c}$.
\item The complement of $R_K$ is a natural example of an computably enumerable set whose
 characteristic sequence has high time bounded Kolmorogov
 complexity for every $n$.
\item $\BPP$ is truth-table reducible to $R_K$.
\item A poly- up-to exponential-size
    advice that has very large $K^{t(n)}$ complexity can be replaced
    by $O(n \log n)$ bit advice and true randomness.
\end{itemize}

\section{Preliminaries}

We remind the reader of some of the definitions we use. Let $M$ be a
Turing machine. For any string $x\in\{0,1\}^*$, the Kolmogorov
complexity of $x$ relative to $M$ is $K_M(x)=\min \{\ |p|\ |\
p\in\{0,1\}^* \;\&\; M(p)=x\}$, where $|p|$ denotes the length of
string $p$. It is well known that for a {\em universal} Turing
machine $U$ and any other machine $M$ there is a constant $c_M$ such
that for all strings $x$, $K_U(x) \le K_M(x) + c_M$. For the rest of
the paper we will fix some universal Turing machine $U$ and we will
measure Kolmogrov complexity relative to that $U$. Thus, we will not
write the subscript $U$ explicitly.

We define $K^t(x)=\min\{\ |p|\ |\ U(p)=x$ and $U(p)$ uses at most
$t(|x|)$ steps$\}$. Unlike traditional computational complexity the
time bound is a function of the length of the output of $U$.

A string $x$ is said to be {\em Kolmogorov-random} if $K(x)\ge |x|$.
The set of Kolmogorov-random strings is denoted by
$R_K=\{x\in\{0,1\}^*\;|\;K(x)\ge |x|\}$. For an integer $n$ and set
$A\subseteq \{0,1\}^*$, $A^{=n} = A\cap \{0,1\}^n$. The following
well known claim can be proven by considering the Kolmogorov
complexity of $|R_K^{=n}|$ (see~\cite{LV}).
\begin{proposition}\label{p-rkdensity}
There is a constant $d$ such that for all $n$, $|R_K^{=n}|\ge 2^{n}/d$.
\end{proposition}

We also use computation with advice. We deviate slightly from the usual
definition of computation with advice in the way how we express and measure
the running time. For an {\em advice function} $\alpha:\N \rightarrow \{0,1\}^*$,
we say that $L\in \Ptime /\alpha$ if there is a Turing machine $M$
such that for every $x\in\{0,1\}^*$, $M(x,\alpha(|x|))$ runs in time polynomial in the
length of $x$ and $M(x,\alpha(|x|))$ accepts iff $x\in L$. We assume that
$M$ has random access to its input so the length of $\alpha(n)$ can grow faster than any polynomial in $n$.
Similarly, we define $\EXP/\alpha$ where we allow the machine $M$ to run in exponential
time in length of $x$ on the input $(x,\alpha(|x|))$. Furthermore, we are interested
not only in Boolean languages (decision problems) but also in functions, so we naturally
extend both definitions also to computation with advice of functions.
Typically we are interested in the amount of advice that we need for inputs of length $n$
so for $f:\N\rightarrow \N$, $C/f$ is the union of all $C/\alpha$
for $\alpha$ satisfying $|\alpha(n)|\le f(n)$.

Let $L$ be a language and $C$ be a language class. We say that
$L\in{\rm i.o.-}C$ if there exists a language $L'\in C$ such that
for infinitely many $n$, $L^{=n} = L'^{=n}$. For a Turing machine
$M$, we say $L\in\ioM/f$ if there is some advice function $\alpha$
with $|\alpha(n)|\leq f(n)$ such that for infinitely many $n$,
$L^{=n}=\{ x\in\Sigma^n\ |\ M(x,\alpha(|x|))$ accepts$\}$.

We say that a set $A$ polynomial-time Turing reduces to a set $B$, if there is an oracle
machine $M$ that on input $x$ runs in polynomial time and with oracle $B$ decides whether
$x\in A$. If $M$ asks its questions {\em non-adaptively}, i.e., each oracle question does not depend
on the answers to the previous oracle questions, we say that $A$ polynomial-time truth-table reduces
to $B$ ($A\ptt B$). Moreover, $A\pdtt B$ if machine $M$ outputs as its answer the disjunction
of the oracle answers. Similarly, $A\pctt B$ for the conjunction of the answers.

\section{High circuit complexity of $R_K$}
In this section we prove that the characteristic sequence of $R_K$
has high circuit complexity almost everywhere. We will first prove
the following lemma.
\begin{lemma}\label{lemma:machines}
For every total Turing machine  $M$ there is a constant $c_M$ such
that $R_K$ is not in $\ioM/2^{n-c_M}$.
\end{lemma}

There is a (non-total) Turing machine $M$ such that $R_K$ is in
$M/n+1$ where the advice is the number of strings in $R_K^{=n}$.
Simply find all the non-random strings of length $n$. This machine
will fail to halt if the advice underestimates the number of random
strings.

\begin{proofof}{Lemma~\ref{lemma:machines}}
Suppose the theorem is false. Fix a total machine $M$. We have that,
$(x, \alpha)
 \in L(M)$ if and only if
$x \in R_K$, for some advice $\alpha$ of length $k\leq 2^{n-c_M}$ and
every $x$ of some large enough length $n$. By padding the advice we
can assume $k=2^{n-c_M}$. We will set $c_M$ later in order to
 get a contradiction.

Let $R_\beta=\{x\in\Sigma^n\ |\ (x,\beta)\in L(M)\}$. By
Proposition~\ref{p-rkdensity} for some constant $d$, $|R_\alpha|\geq
2^n/d$ so we know that if $|R_\beta|<2^n/d$ then $\beta\neq\alpha$.
We call $\beta$ good if $|R_\beta|\geq 2^n/d$.

Fix a good $\beta$ and choose $x_1,\dots,x_m$ at random. The
probability that all the $x_i$ are not in $R_\beta$ is at most
$(1-1/d)^m< 2^{-m/d}$. There are $2^k$ advice strings $\beta$ of length $k$ so if
$2^{-m/d}\leq 2^{-k}$ then there is a sequence $x_1,\ldots,x_m$ such
that for every good $\beta$ of length $k$ there is an $i$ such that
$x_i\in R_\beta$.

We can computably search all such sequences so let $x_1,\ldots,x_m$
be the lexicographically least sequence such that for each good
$\beta$ of length $k$, there is some $x_i\in R_\beta$. This also
means $x_i\in R_\alpha$ for some $i$ so for one of the $x_i$ we have
$K(x_i)\geq n$.

Fix $m=2^{n-a}$ for a constant $a$ to be chosen later.

We can describe $x_i$ by $n-a+b\log a$ bits for some constant $b$:
$n-a$ bits to describe $i$, $O(\log a)$ bits to recover $n$ and a
constant number of additional bits to describe $k$, $M$, $d$ and the
algorithm above for finding $x_1,\ldots,x_m$. If we pick $a$ such
that $a>b\log a$ we contradict the fact that $K(x_i)\geq n$.

If we pick $c_M\geq a+\log d$ we then have $2^{n-a}\geq 2^{n-c_M}d$,
$m>kd$ and $2^{-m/d}\leq 2^{-k}$ completing our contradiction.
\end{proofof}

In order to get our statement about time bounded advice classes we
instantiate Lemma~\ref{lemma:machines} with universal machines $U_t$
that run in time $t$, use the first part of their advice,  in prefix
free form, as a code for a machine that runs in time $t$ and has the
second part of the advice for $U_t$ as its advice. The following is
a direct consequence of Lemma~\ref{lemma:machines}.
\begin{lemma}\label{lemma:universal}
For every computable time bound $t$ and universal advice machine
$U_t$ there is a constant $c_t$ such that $R_K$ is not in
$\ioU_t/2^{n-c_t}$.
\end{lemma}

We are now ready to prove the main theorem from this section.

\begin{theorem}\label{t-RK}
For every computable time bound $t$ there is a constant $d_t$ such
that $R_K$ is not in $\ioDTIME(t)/2^{n-d_t}$.
\end{theorem}
\begin{proof}
Suppose the theorem is false, that is there is a time bound $t$ such
that for every $d$ there is a machine $M_d$ that runs in time $t$
such that $R_k \in \ioM_d/2^{n-d}$. Set $t'= t\log t$ and let
$c_{t'}$ be the constant that comes out of
Lemma~\ref{lemma:universal} when instantiated with time bound $t'$.
Set $d=c_{t'}+1$ and let the code of machine $M_d$ from the
(false) assumption have size $e$. So we have that $R_k \in
\ioM_d/2^{n-d}$. This in turn implies that $R_K \in
\ioU_{t'}/2^{n-d}+e+2\log e$, which implies that $R_K \in
\ioU_{t'}/2^{n-c_{t'}}$ a contradiction with
Lemma~\ref{lemma:universal}. The last step is true because the
universal machine running for at most  time $t'=t \log t$, can simulate $M_d$,
who runs in time $t$.

\end{proof}

As an immediate corollary we get an alternative, more natural
candidate for Barzdin's computably enumerable set that has high
resource bounded Kolomorov complexity, namely the set of
compressible strings.

\begin{corollary}
For every computable time bound $t$ there is a constant $c$ such
that $K^t(\overline{R_k}(1:n)\mathrel{|}n) \geq n/c$
\end{corollary}

Barzdin~\cite{Barzdin} also showed that this lower bound is optimal.
That is the dependence of $c$ on the time bound $t$ is needed for
the characteristic sequence of every r.e. set. Hence this depence is
also necessary in our Theorem~\ref{t-RK}.

% \begin{proof}
%   \cite{NW:1994}'s pseudo-random generator can be used to derandomise $\BPP$, when given a $(2^{\epsilon n}, 2^{\epsilon n})$-hard function $f$. The generator is a function $G_f : \{ 0, 1 \}^s \to \{ 0, 1 \}^n$  which takes a seed of length $s = O(\log n)$, and queries $f$ on inputs $x_1, \ldots, x_n$ also of length $O(\log n)$, constructed by selecting specific bits from the seed. So in order to compute $G_f$, it is only necessary to have the values of $f$ for every input of length up to $s$, and these values are $poly(s)$-time truth-table reducible to $R_K$, and $poly(n)$-time truth-table reducible to $\mathrm{SPARSE}(R_K)$.
% \end{proof}

\section{$\BPP$ truth-table reduces to $R_k$}

In this section we investigate what languages are reducible to
$R_k$. We start with the following theorem which one can prove using
nowadays standard derandomization techniques.

\begin{theorem}\label{thm:derandomization}
Let $\alpha:\{0\}^* \rightarrow \{0,1\}^*$ be a length preserving
function and $\delta>0$ be a constant. If $\alpha(0^n)
\not\in\ioEXP/n^\delta$ then  for every $A\in\BPP$ there exists
$d>0$ such that $A\in \P/\alpha(0^{n^d})$.
\end{theorem}

\begin{proof}
$\alpha(0^n)\not\in \ioEXP/n^\delta$ implies that when $\alpha(0^n)$
is interpreted as a truth-table of a function
$f_{\alpha(0^n)}:\{0,1\}^{\log n} \rightarrow \{0,1\}$,
$f_{\alpha(0^n)}$ does not have boolean circuits of size
$n^{\delta/3}$ for all $n$ large enough. It is known that such a
function can be used to build the Impagliazzo-Wigderson pseudorandom
generator \cite{IW97} which can be used to derandomize boolean
circuits of size $n^{\delta'}$ for some $\delta'>0$ (see \cite{IW97,
KvM99, ABKvMR02}). Hence, bounded-error probabilistic computation
running in time $n^\ell$ can be derandomized in polynomial time
given access to $\alpha(0^{n^{2\ell/\delta'}})$.
\end{proof}

%% Note one could also use the weaker assumption that
%% $\alpha(0^n)\not\in\ioP/n^\delta$ in this proposition.
From Theorem \ref{t-RK} and the above Theorem we obtain the
following corollary.

\begin{corollary}
 $\BPP \le^p_{tt} R_K$.
\end{corollary}

\begin{proof}
Let $\alpha(0^n)$ be the truth-table of $R_K$ on strings of length
$\lfloor \log n \rfloor$ padded by zeros to the length of $n$. By
Theorem \ref{t-RK}, $\alpha(0^n)\not\in \ioEXP/(n/c)$ for some
$c>0$. Consider any $A\in \BPP$. By
Theorem~\ref{thm:derandomization} for some $d$, $A\in
P/\alpha(0^{n^d})$. The claim follows by noting that a truth-table
reduction to $R_k$ may query the membership of all the strings of
length $\lfloor \log n^d \rfloor$ to construct $\alpha(0^{n^d})$ and
then run the $P/\alpha(0^{n^d})$ algorithm for $A$.
\end{proof}

Our goal would be to show that using  $R_K$ as a source of
randomness is the only way  to make use of it. Ideally we would
like to show that any recursive set that is truth-table reducible to
$R_K$ must be in $\BPP$. We fall short of such a goal. However we
can show the following claim.

\begin{theorem}\label{t-op1}
Let $\alpha:\{0\}^* \rightarrow \{0,1\}^*$ be a length preserving
function and $c>0$ be a constant. If $\alpha(0^n) \not\in\ioEXP/n -
c\log n$ then for every $A\in\EXP$ if $A\in \PTIME/\alpha(0^{n^d})$
for some $d>0$ then $A\in\BPP/O(n\log n)$.
\end{theorem}

This theorem says that Kolmogorov random advice of polynomial size
can be replaced by almost linear size advice and true randomness. We
come short of proving a converse of the above corollary in two
respects. First, the advice is supposed to model the initial segment
of the characteristic sequence of $R_K$ which the truth-table can
access. However, by providing only polynomial size advice we
restrict the hypothetical truth-table reduction to query strings of
only logarithmic length. Second, the randomness that we require from
the initial segment is much stronger than what one can prove and
what is in fact true for the initial segment of the characteristic
sequence of $R_K$. One can deal with the first issue as is shown by
Theorem~\ref{t-op3} but we do not know how to deal with the second
one.

%??? Maybe add more conclusions from $\NP \in \BPP$. ???

% Lance: I've taken this out unless Michal wants to check/extend it
%The
%theorem can be further extended to the case of exponentially long
%advice $\alpha(0^{2^n})$ with random access. Under the assumption
%that $\alpha(0^n)\not\in\ioEXP/n - c\log \log n$ the same
%conclusions hold. ??? Double-check ??? Michal can you extend this?

\begin{proof}
Let $M$ be a polynomial time Turing machine and $A\in \EXP$ be a set
such that $A(x) = M(x, \alpha(|x|^d))$. We claim that for all $n$
large enough there is a non-negligible fraction of advice strings
$r$ of size $n^d$ that could be used in place of $\alpha(n^d)$ more
precisely:
$$\Pr_{r\in\{0,1\}^{n^d}}[\forall x, x\in A \iff M(x,r) = 1] >
\frac{1}{n^{cd}}.$$
To prove the claim consider the set $G=\{r\in\{0,1\}^{n^d};\;\forall
x\in\{0,1\}^n, x\in A \iff M(x,r) = 1\}$. Clearly, $G\in\EXP$ and
$\alpha(0^{n^d})\in G$. If $|G^{=n^d}|\le 2^{n^d}/n^{cd}$ then
$\alpha(0^{n^d})$ can be computed in exponential time from its index
in the set $G^{=n^d}$ of length $n^d- cd \log n$. Since
$\alpha(0^{n^d}) \not\in\ioEXP/n^d - cd\log n$ this cannot happen
infinitely often.

% If this wasn't the case for all sufficiently large $n$, we would infinitely often only
% need $n^c - c \log n$  bits of advice to describe $\alpha(n^c)$ among the list of ``good $R$s''.
% But since $A \in EXP$, this list of good $R$s is obtainable in $\EXP$, this contradicts
% the hardness of $\alpha$.

Now we present an algorithm that on input $x$ {\em samples} from $G$
using only $O(n \log n)$ bits of advice (in fact $O(\log n)$ entries
from the truth table of $A$) and outputs $A(x)$ with high
probability. Consider the following algorithm:
\begin{enumerate}
\item Given an input $x$ of length $n$, and an advice
    string $x_1,A(x_1), . . . , x_k,A(x_k)$,
\item sample at most $2n^{cd}$ strings of length $n^d$
    until the first string $r$ is found such that $M(x_i,r) =
    A(x_i)$ for all $i\in\{1,\dots,k\}$.
\item If we find $r$ consistent with the advice then output
    $M(x,r)$ otherwise output 0.
\end{enumerate}

For all $n$ large enough the probability that the second step does
not find  $r$ compatible with the advice is upper-bounded by the
probability that we do not sample any string from $G$ which is at
most $(1-\frac{1}{n^{cd}})^{2n^{cd}} < e^{-2} < 1/6$.

It suffices to show that we can find an advice sequence such that
for at least $5/6$-fraction of the $r$'s compatible with the advice
$M(x,r)=A(x)$. For given $n$, we will find the advice by prnning
iteratively the set of bad random strings $B=\{0,1\}^{n^d} \setminus
G$. Let $i=0,1,\dots,2cd \log_{6/5} n$. Set $B_0=B$. If there is a
string $x\in\{0,1\}^n$ such that for at least $1/6$ of $r\in B_i$,
$M(x,r)\not= A(x)$, then set $x_{i+1}=x$ and $B_{i+1}=B_i \cap
\{r\in\{0,1\}^{n^d}\mathrel{|}M(x_{i+1},r)=A(x_{i+1})\}$. If there is no such
string $x$ then stop and the $x_i$'s obtained so far will form our
advice. Notice, if we stop for some $i<2cd \log_{6/5} n$ then for
all $x\in\{0,1\}^n$, $\Pr_{r\in\B_i}[M(x,r)\not=A(x)]<1/6$. Hence,
any $r$ found by the algorithm to be compatible with the advice will
give the correct answer for a given input with probability at least
$5/6$. On the other hand, if we stop building the advice at $i=2cd
\log_{6/5} n$ then $|B_{3cd \log_{6/5} n}|\le 2^{n^d} \cdot
(5/6)^{2cd \log_{6/5} n} \le |G^{=n^d}|/n^{cd}$. Hence, any string
$r$ found by the algorithm to be compatible with the advice
$x_1,A(x_1), . . . , x_i,A(x_i)$ will come from $G$ with
good  probability, i.e., with probability $>5/6$ for $n$
large enough.
%
% Step 2 will always find such an R with high probability, since there is a nonnegligible
% fraction of such Rs. We build the advice string in the following way:
% Call an R âgood up to xkâ if M(xi,R) = Axi for all i ( k; xk+1 is chosen
% so that at least a 1/3 fraction of the âgoodâ Rs is not good for xk+1, i.e., at
% least 1/3 of the Rs good up to xk have M(xk+1,R) )= A(xk+1). If no such xk+1
% exists, then the advice string we want is x1,A(x1), . . . , xk,A(xk).
% Notice that such process ends after O(log n) steps, since by each step we cut
% out a constant fraction of the remaining Rs. When the process is over, every x
% can be predicted with probability at least 2/3 using the Rs good up to xk.
\end{proof}

The following theorem can be established by a similar argument. It
again relies on the fact that a polynomially large fraction of all
advice strings of length $2^{n^d}$ must work well as an advice. By a
pruning procedure similar to the proof of Theorem \ref{t-op1} we can
avoid bad advice. In the $\BPP$ algorithm one does not have to
explicitly guess the whole advice but only the part relevant to the
pruning advice and to the current input.

\begin{theorem}\label{t-op3}
Let $\alpha:\{0\}^* \rightarrow \{0,1\}^*$ be a length preserving
function and $c>0$ be a constant. If $\alpha(0^n) \not\in\ioEXP/n -
c\log \log n$ then for every $A\in\EXP$ if $A\in
\PTIME/\alpha(0^{2^{n^d}})$ for some $d>0$ then $A\in\BPP/O(n\log
n)$.
\end{theorem}

We show next that if the set $A$ has some suitable properties
we can dispense with the linear advice all together and
replace it with only random bits.  Thus for example if
$\SAT\in\PTIME/\alpha(0^n)$ for some
computationally hard  advice $\alpha(0^n)$ then $\SAT\in\BPP$.

\begin{theorem}\label{t-op2}
Let $\alpha:\{0\}^* \rightarrow \{0,1\}^*$ be a length preserving
function and $c>0$ be a constant such that $\alpha(0^n) \not\in\ioEXP/n -
c\log n$. Let $A$ be paddable and polynomial-time many-one-complete for
a class $\C \in \{\NP,\PTIME^{\#\PTIME}, \PSPACE, \EXP\}$.
If $A\in \PTIME/\alpha(0^{n^d})$ for some $d>0$ then $A\in \BPP$
(and hence $C \subseteq \BPP$).
\end{theorem}

To prove the theorem we will need the notion of
instance checkers. We use the definition of Trevisan and Vadhan
\cite{TV02}.

%% Note the following definition and theorem is verbatim from Trevisan and Vadhan.

\begin{definition}
An {\em instance checker} $C$ for a boolean function $f$ is a
polynomial-time probabilistic oracle machine whose output is in
$\{0, 1, {\mathrm{fail}}\}$ such that
\begin{itemize}
\item for all inputs $x$, $\Pr[C^f(x) = f(x)] = 1$, and
\item for all inputs $x$, and all oracles $f'$,
    $\Pr[C^{f'}(x) \not\in \{f(x), {\mathrm{fail}}\}] \le 1/4$.
\end{itemize}
\end{definition}

It is immediate that by linearly many repetitions and taking the
majority answer one can reduce the error of an instance checker to
$2^{-n}$. Vadhan and Trevisan also state the following claim:

\begin{theorem}[\cite{BFL},\cite{LFKN,Shamir92}] Every problem that is
complete for $\EXP$, $\PSPACE$ or $\PTIME^{\#\PTIME}$ has an
instance checker. Moreover, there are $\EXP$-complete problems,
$\PSPACE$-complete problems, and $\PTIME^{\#\PTIME}$-complete
problems for which the instance checker $C$ only makes oracle
queries of length exactly $\ell(n)$ on inputs of length $n$ for some
polynomial $\ell(n)$.
\end{theorem}

However, it is not known whether $\NP$ has instance checkers.

\begin{proofof}{Theorem~\ref{t-op2}}
To prove the claim for $\PTIME^{\#\PTIME}$-,
$\PSPACE$- and $\EXP$-complete problems we use the instance
checkers. We use the same notation as in the proof of Theorem \ref{t-op1}, i.e.,
$M$ is a Turing machine such that $A(x) = M(x, \alpha(|x|^d))$
and the set of good advice is $G=\{r\in\{0,1\}^{n^d};\;\forall
x\in\{0,1\}^n, x\in A \iff M(x,r) = 1\}$. We know from the previous
proof that $|G^{=n^d}|\ge 2^{n^d}/n^{cd}$ because  $\alpha(0^n) \not\in\ioEXP/n - c\log n$.

Let $C$ be the instance checker for $A$ which on input of
length $n$ asks oracle queries of length only $\ell(n)$ and makes
error on a wrong oracle at most $2^{-n}$. The following algorithm is
a bounded error polynomial time algorithm for $A$:
\begin{enumerate}
\item On input $x$ of length $n$, repeat $2n^{cd}$ times
\begin{enumerate}
\item Pick a random string $r$ of length $(\ell(n))^d$.
\item Run the instance checker $C$ on input $x$ and
    answer each of his oracle queries $y$ by $M(y,r)$.
\item If $C$ outputs $\mathrm{fail}$ continue with
    another iteration otherwise output the output of $C$.
\end{enumerate}
\item Output 0.
\end{enumerate}

Clearly, if we sample $r\in G$ then the instance checker will
provide a correct answer and we stop. The algorithm can produce a
wrong answer either if the instance checker always fails (so we
never sample $r\in G$ during the iterations) or if the instance
checker gives a wrong answer. Probability of not sampling good $r$
is at most $1/6$. The probability of getting a wrong answer from the
instance checker in any of the iterations is at most $2n^{cd}/2^n$.
Thus the algorithm provides the correct answer with probability at least
$2/3$.

To prove the claim for $\NP$-complete languages we
show it for  the canonical example of SAT. The following algorithm
solves SAT correctly with probability at least $5/6$:
\begin{enumerate}
\item On input $\phi$ of length $n$, repeat $2n^{cd}$ times
\begin{enumerate}
\item Pick a random string $r$ of length $n^d$.
\item If $M(\phi,r)=1$ then use the self-reducibility of
    SAT to find a presumably satisfying assignment $a$ of
    $\phi$ while asking queries $\psi$ of size $n$ and
    answering them according to $M(\psi,r)$. If the
    assignment $a$ indeed satisfies $\phi$ then output 1
    otherwise continue with another iteration.
\end{enumerate}
\item Output 0.
\end{enumerate}

Clearly, if $\phi$ is satisfiable we will answer 1 with probability
at least $5/6$. If $\phi$ is not satisfiable we will always answer
0.
\end{proofof}

\section{Open Problems}

We have shown that the set $R_K$ cannot be compressed using a
computable algorithm and used this fact to reduce $\BPP$
non-adaptively to $R_K$. We conjecture that every computable set
that non-adaptively reduces in polynomial-time to $R_K$ sits in
$\BPP$ and have shown a number of partial results in that
directions.

The classification of languages that polynomial-time adaptively
reduce to $R_K$ also remains open. Can we characterize $\PSPACE$
this way?

\bibliographystyle{alpha}
\bibliography{bppr}

\end{document}